
\input harvmac.tex
\Title{\vbox{\baselineskip12pt
\hbox{USC-93-020}}}
{\vbox{\centerline{Lattice regularization of}
\vskip2pt\centerline{massive and massless }
\vskip2pt\centerline{integrable  field theories}}}

\centerline{N.Yu Reshetikhin$^\dagger$, H.Saleur\foot{Packard Fellow}
$^\spadesuit$}
\vskip2pt
\centerline{$^\dagger$Department of Mathematics, Berkeley University}
\centerline{Berkeley CA 94720}
\vskip2pt
\centerline{$^\spadesuit$Department of Physics and Department of Mathematics}
\centerline{University of southern California}
\centerline{Los Angeles CA 90089}
\vskip.3in

We show that  integrable vertex and RSOS models with trigonometric
Boltzmann weights and appropriate inhomogeneities provide a
 convenient lattice regularization
 for   massive field theories and for  the recently studied  massless
field theories that interpolate between two non trivial conformal
 field theories.  Massive and massless S matrices are computed from the
lattice Bethe ansatz.

\newsec{Introduction}

The   perturbation of  unitary minimal models ${\cal M}_{t}$ of
 conformal field theory with central charge
\eqn\cuv{c=1-{6\over t(t-1)},}
by the $\phi_{13}$ operator of conformal weight
\eqn\hhh{h={t-2\over t},}
is known to give an integrable 1+1 quantum field theory
\ref\SZamo{A.B.Zamolodchikov, Int. Jour. Mod. Phys. A3 (1988), 746.}:
 ${\cal M}A_{t}^\pm$. The infra red (IR) properties of this theory depend
 on the sign of the coupling. Writing the hamiltonian as
\eqn\act{H_{{\cal M}A_{t}}=H_{{\cal M}_{t}}+g\int
\phi_{13}(x)\bar{\phi}_{13}(x)
dx,}
the case $g<0$ gives the  massive integrable field theory ${\cal MA}_t^-$ with
 a trivial IR fixed point while the case
 $g>0$ is "massless" and gives ${\cal MA}_t^+$ with an IR fixed point
that coincides with
${\cal M}_{t-1}$\SZamo\ \ref\CL{J.Cardy, A.Ludwig, Nucl. Phys. B287
(1987) 687.}.

The conformal field theory ${\cal M}_{t}$ is well known to describe the
continuum limit of
the $(t-2)^{th}$ critical Ising model \ref\SZamoI{A.B.Zamolodchikov,
Sov. J. Nucl. Phys. B287 (1987) 687.}, a critical point where  in the
 Landau Ginzburg picture $t-2$ minima of the potential coalesce. The massive
 flow describes the flow towards a phase where these minima separate and are
 degenerate and where spontaneous symmetry breaking occurs, while the massless
 one corresponds to one minima decoupling so in the IR the model is a
multicritical Ising model with one order of criticality less.

It is known that minimal models ${\cal M}_t$ of conformal field theory can be
obtained by twisting of boundary conditions and then reduction of the
Gaussian model with $c=1$ \ref\DFetc{V.Dotsenko, V.Fateev,
 Nucl. Phys. B251 (1985) 691; G.Felder, Nucl. Phys. B317 (1989) 215.}.
In a similar way, the massive regime of the
 perturbed theory with action \act\
 can be described as a  reduction of the sine-Gordon model
 \ref\RS{N.Yu Reshetikhin, F.Smirnov, Comm. math. Phys. 131 (1990) 157.} at
 coupling ${\beta^2\over 8\pi}={t-1\over t}$. Due to twisting, the coupling
constant in front of the cosine term scales as $G\propto(-g)^{1/2}$.
The soliton mass $m$ scales as $m\propto(-g)^{{t\over 4}}$.

Formally the same correspondence between the two theories should also hold
in the massless regime. In that case however the sine-Gordon coupling
 constant  $G$ turns out to be purely imaginary due to the above scaling.
 This is a regime that has not been much studied so far, and which presents
subtleties \ref\Fetc{P.Fendley, H.Saleur, Al.B.Zamolodchikov, "Massless
flows between minimal models", preprints USC-93/003 and 93/004.}. Notice of
 course that the rather
 violent non unitarity of this sine-Gordon model does not prevent the subset
 of observables corresponding to the minimal model to be unitary.

Perturbations of coset models  $su(2)_s\otimes su(2)_{t-s-2}/su(2)_{t-2}$
 ($s=1$ above) by the operator  of  dimension $h={t-2\over t}$ also give
 integrable models. We will denote them
 ${\cal M}A^\pm_{s,t}$ depending on the sign of the coupling constant in front
of the perturbation.
 These models can also be obtained by reduction of some other
 integrable models: the  higher spin or  fractional supersymmetric
 generalizations  of the sine-Gordon
 model  \ref\BL{D.Bernard, A.Leclair, Phys. Lett. B247 (1990) 309.}
\ref\Zamo{Al.B.Zamolodchikov, Nucl. Phys. B366 (1991) 122.}.

 Further
 generalizations to coset models of type $G_k\times G_l/G_{k+l}$ are also
 possible. For
each such model there is a natural integrable direction of deformation
\ref\EY{T.Eguchi, S.K.Yang, Phys. Lett. B224 (1989) 373.} with the perturbing
field of dimension $1-{h^v\over k+l+h^v}$ (where $h^v$ is the dual Coxeter
number) leading to two different theories depending on the sign
of the coupling constant. The role of the sine-Gordon model is then
played by its Toda generalizations and further fractional supersymmetric
generalizations \ref\ABL{C.Ahn, D.Bernard, A.Leclair, Nucl. Phys.
 B346 (1990) 409.}.

We show in this paper how inhomogeneous vertex or RSOS  models with
trigonometric Boltzmann weights and appropriate inhomogeneities
provide, in their scaling limit, a general lattice
 regularization of the above massive and massless field theories.
 This allows us in particular to put the
 massless scattering of
 \ref\ZamoI{Al.B.Zamolodchikov, Nucl. Phys. B358 (1991) 524.}
 on firmer grounds, and to prove a conjecture in \Fetc\ . In a way our approach
is parallel to the one developed on \ref\FR{L.D.Faddeev, N.Yu Reshetikhin,
 Ann. Phys. 167 (1986) 227.} for non linear sigma models. It is interesting
that the lattice
version of the sine-Gordon model which was studied in \ref\IK{A.G.Izergin,
 V.E.Korepin, Lett. Math. Phys. 5 (1981) 199.} turns out to be of the
same nature. It was shown in \ref\FV{L.D.Faddeev, A.Yu Volkov, Theor.
 Math. Phys. 92 (1992) 207.} that it can also be formulated as a
 model on the inhomogeneous lattice where at each vertex there
is a cyclic representation of
$U_qsl(2)$.

Although the method is general, we discuss mainly the $su(2)$ case. Appropriate
extensions to other groups are discussed in the last section. In section 2 we
 recall
necessary results about the Bethe ansatz for inhomogeneous vertex and RSOS
 models. In section 3 we discuss the thermodynamic limit. In section 4 we
consider
the
scaling limit. In section 5 we study the S
 matrix
in the massless case and prove the conjecture of \Fetc\ . In section 6 we
 discuss
some examples of the method applied to other groups.

Lattice regularizations of masssive field theories
 which  are based on inhomogeneous  integrable
 models with trigonometric  Boltzmann weights have some advantages compared
with
the elliptic ones
 \ref\L{M.Luscher, Nucl. Phys. B117 (1976) 475.}.
One is that in this case the lattice and continuum theory have the same
symmetry.
Another  is that we know far more about the spectrum of trigonometric
transfer matrices than about the elliptic ones. Finally for Lie algebras other
than $su(n)$ we do not expect elliptic  vertex models
\ref\BD{A.A.Belavin, V.G.Drinfeld, Soviet Scientific Reviews C4 (1984) 93.}.

\newsec{Inhomogeneous lattice models of $XXZ$ type }

\subsec{Vertex models}

Consider a one dimensional lattice with N vertices formed by the intersection
of N columns and one row. We assume the vertices are numbered from left to
 right. A $U_qsl(2)$ or generalized 6-vertex model on such lattice can be
described
as follows. Fix spins $s_j$, $j=1,\ldots,N$ (in units where the fundamental
 has spin one) and an additional spin $s$. With each vertical edge we associate
the space of states $C^{s_j+1}$ and a complex number $u_j$. With each
 horizontal edge we associate $C^{s+1}$ and a complex number $u$.

\bigskip
\noindent
\centerline{\hbox{\rlap{\raise30pt\hbox{$\hskip.75cm u_1\ldots\hskip.8cm
u_j\ldots$}}
\rlap{\raise10pt\hbox{$\hskip1.cm\Big|\hskip1.cm\Big|\hskip1.cm\Big|
\hskip1.cm\Big|$}}
\rlap{\lower10pt\hbox{$\hskip.88cm\Big|\hskip1.cm\Big|\hskip1.cm\Big|
\hskip1.cm\Big|$}}
\rlap{\lower30pt\hbox{$\hskip.6cm s_1\ldots\hskip1.3cm s_j\ldots$}}
$\hskip-.4cm u$    -------------------------------------------$
\hskip.5cm s$}}

\bigskip
\noindent
 With each vertex
we associate a system of Boltzmann weights which are given by R-matrices
acting in the tensor product of corresponding spaces.
 In each
space $C^{s+1}$ we enumerate the basis by $a=1,\ldots,s+1$. Fix an element of
the basis in
each space of states corresponding to the boundary edges, say $a,a_1,
\ldots,a_N,b,b_1,\ldots,b_N$.
The partition function of the $U_qsl(2)$ model with the parameters
 and boundary conditions just described  has the following form
\eqn\Ki{T^s(u|\{u_j\}|\{s_j\})^{a,a_1\ldots a_N}_{b,b_1\ldots b_N}=
\sum_{c_1,\ldots,c_{N-1}}R^{ss_1}(u-u_1)^{aa_1}_{c_1b_1}\ldots
R^{ss_N}(u-u_N)^{c_{N-1}a_N}_{bb_N}.}
Here $R^{ss_j}$ is the $U_qsl(2)$ R-matrix described in
\ref\KuRS{P.Kulish, N. Reshetikhin, E.Sklyanin, Lett. Math. Phys. 5 (1981)
393.}
and \ref\J{M.Jimbo,
 Lett. Math. Phys. 10 (1985) 163.}.

{}From this partition function with open boundary conditions on one row we can
 construct
corresponding partition functions on rectangular lattices with various boundary
conditions as traces of products of \Ki\ .

Since the R-matrices satisfy the Yang Baxter equation we have commutativity
 of traces
\eqn\Kii{\left[t^s(u|\{u_j\}|\{s_j\}),t^{s'}(v|\{u_j\}|\{s_j\})\right]=0.}
Here the transfer matrices $t^s(u|\{u_j\}|\{s_j\})$ act in the tensor product
${\cal H}^{vertex}_{s_1,\ldots,s_N}=C^{s_1+1}\otimes\ldots\otimes C^{s_N+1}$
and they are given by
traces of \Ki\ with respect to "horizontal spaces"
\eqn\Kiii{t^s(u|\{u_j\}|\{s_j\})^{a_1\ldots a_N}_{b_1\ldots b_N}=
\sum_{a=1}^{s+1} d_a T^s(u|\{u_j\}|\{s_j\})^{a, a_1\ldots a_N}_{a,
b_1\ldots b_N},}
where $d_a=\exp[\kappa (a-1-s/2)]$ for some complex number $\kappa$. These
transfer matrices commute with the total third component $s^z$ of the spin.
The space
 ${\cal H}^{vertex}_{s_1,\ldots,s_N}$ we may call vertical space.
We can thus regard \Kii\ as a generating function for a family of quantum
integrable systems. The commuting family  can be diagonalized simultaneously.
The eigenvectors of $t^s(u|\{u_j\}|\{s_j\})$ are parametrized by solutions
$\{\alpha_k\}$
to the Bethe equations
\eqn\ba{\prod_{j=1}^{j=N}\exp\left[ip_{s_j}(\alpha_k-2iu_j)
\right]\prod_{l=1}^{l=M}\exp\left[i\phi(\alpha_k-\alpha_l)\right]=-1,}
where $M=(\sum_js_j-s^z)/2$ is an integer. The functions $p_s$  and $\phi$
are given respectively by
\eqn\kri{\exp ip_s(\alpha)={\sinh{1\over 2}(\alpha-si\gamma)\over
\sinh{1\over 2}(\alpha+si\gamma)},}
and
\eqn\krii{\exp i\phi(\alpha)={\sinh{1\over 2}(\alpha+2i\gamma)\over
\sinh{1\over 2}(\alpha-2i\gamma)}.}
With these conventions the dependence on heterogeneities in the vertical spaces
appears only in the Bethe equations. Eigenvalues depend on the horizontal
parameters
$u,s$ and the numbers $\alpha_k$ as in the homogeneous case. Complete
expressions,
which  involve a sum of $s+1$ terms, can be found in
 \ref\KR{A.N.Kirillov, N.Yu Reshetikhin,
J.Phys. A 20 (1987) 1565, 1587.}.
The dominant term in the thermodynamic limit
reads (with proper normalization)
%
\eqn\eig{\Lambda_N^{ss'}(u)\approx\prod_k\exp\left\{-i[p_{s'}(\alpha_k+2iu)-\pi]
\right\},\ N\rightarrow\infty.}
For more details and references see for instance \KR\ .

\subsec{RSOS models}

If $q$ is a root of unity, $q=\exp(i\pi/t)$ where $t$ is an integer, we can
also construct the Restricted  Solid On Solid (RSOS)  models corresponding
to the quantum group $U_qsl(2)$. States in such a model are associated with
faces of the lattice, and they take values $l_j=1,\ldots,t-1$. The adjacency
conditions between  neighbouring faces depend on the spin associated with the
edge between them \ref\JMO{M.Jimbo, T.Miwa, M.Okado, Nucl. Phys. B300 (1988)
 74.}\ref\ABF{G.E.Andrews, R.J.Baxter, P.J.Forrester, J.Stat. Phys.
35 (1984) 193.}. The Boltzmann weights are associated with vertices of the
 lattice and they depend on the states of the neighbouring four faces. These
 weights are related to the vertex R-matrices via 6j calculus
  \ref\P{V.Pasquier, Comm. Math. Phys. 118 (1988) 355.}\ref\KRI{A.N.Kirillov,
 N.Yu Reshetikhin, in "Infinite-dimensional Lie algebras and groups",
 ed. V.G.Kac, World Scientific, Singapore (1989).}.

Fix spins $s_1,\ldots,s_N$. The corresponding vertical space of states in the
RSOS model, ${\cal H}^{RSOS}_{s_1,\ldots,s_N}$ with periodic boundary
conditions,
is best
 described by its basis
\eqn\kriii{\{l_1,\ldots,l_N\},
\ (l_i-l_{i+1}+s_i)/2\in\{0,\ldots,s_i\},\ s_i<l_i+l_{i+1}<2t-s_i.}

According to \ABF\ ,
\ref\BR{V.V.Bazhanov, N.Yu Reshetikhin, J.Mod. Phys. A4 (1989) 115. } the
eigenvectors of the RSOS transfer matrix are obtained from the Bethe equations
\eqn\barsos{\prod_{j=1}^{j=N}\exp\left[ip_{s_j}(\alpha_k-2iu_j)
\right]\prod_{l=1}^{l=M}\exp\left[i\phi(\alpha_k-\alpha_l)\right]=-\omega^2,}
where we now have the constraint $M=\sum_j s_j/2$ and $\omega^t=-(-)^My$
($\hbox{Im}\omega\neq 0$) and
$y$ is the eigenvalue of the operator given by
$$
Y_{l_1,\ldots,l_N}^{l'_1,\ldots,l'_N}=\prod_{i=1}^{i=N}\delta(l_i,t-l'_i)
$$
that commutes with the RSOS transfer matrix. Eigenvalues of the RSOS transfer
matrix are given by a sum of $s+1$ terms as in the vertex case, each term
being however weighed by different powers of $\omega$. For complete
expressions see \BR\ . These eigenvalues are a subset of the eigenvalues
of the vertex model with $\omega=\exp\kappa$ \Kiii\ and in the spin
$s^z=0$ sector \ref\PS{V.Pasquier, H.Saleur, Nucl. Phys. B330 (1990) 523.}.

\subsec{Arrays}

We shall consider further only two choices of parameters $s_1,\ldots,s_N$
and $u_1,\ldots,u_N$:

\noindent Case I: we choose $s_1=\ldots=s_N=s$ and
$-u_{2j}=u_{2j-1}=i\Lambda/2$
for $j=1,\ldots,N/2$.
\smallskip
\noindent Case II: we choose $s_{2j-1}=s_1,s_{2j}=s_2$
and  $-u_{2j}=u_{2j-1}=i\Lambda/2$ for $j=1,\ldots,N/2$.

 In both cases we assume $N$ to be even and  we  consider transfer matrices
 as generating functions for
hamiltonians of quantum spin models (of generalized XXZ or RSOS type).
 For hamiltonians we choose
\eqn\hami{H_I=-{1\over t}\left.{d\over du}\ln\left[t^s(i\Lambda/2+u)
(t^s(-i\Lambda/2-u))^{-1}\right]\right|_{u=0},}
and
\eqn\hamii{H_{II}=-{1\over t}\left.{d\over du}\ln\left[t^{s_1}(i\Lambda/2+u)
(t^{s_2}(-i\Lambda/2-u))^{-1}\right]\right|_{u=0},}
so we have a total of four possible cases.

In all cases the models consist of an even and odd sublattice and hamiltonians
commute with the total translation operator ${\cal T}: j\rightarrow j+2$.
Hamiltonians
are local and can be written
\eqn\hamiii{H=H_{+}+H_{-},}
with
\eqn\hai{H_{+}(\Lambda)=-{1\over t}\left[\sum_{j\hbox{ odd}}\dot{R}_{j+1,j}
(i\Lambda)+
\sum_{j\hbox{ even}}
\left(R_{j+2,j+1}(i\Lambda)\right)^{-1}\dot{R}_{j+2,j}(0)R_{j+2,j+1}(i\Lambda)\right],}
and
\eqn\haii{H_{-}(\Lambda)=-{1\over t}\left[\sum_{j\hbox{ even}}\dot{R}_{j-1,j}
(-i\Lambda)+
\sum_{j\hbox{ odd}}
R_{j-2,j-1}(-i\Lambda)\dot{R}_{j-2,j}(0)\left(R_{j-2,j-1}(-i\Lambda)\right)^{-1}\right]
,}
where dots stand for ${d\over du}$ and
 R-matrices implicitely carry spin upper indices in agreement
with the arrays I and II, and correspond to XXZ type or RSOS models.

In terms of Bethe ansatz solutions eigenenergies of these hamiltonians read
 from \eig\
\eqn\ener{E_I=-{2\over t}\sum_k \dot{p}_{s}(\alpha_k+\Lambda)+
\dot{p}_{s}(\alpha_k-\Lambda)}
and
\eqn\enerI{E_{II}=-{2\over t}\sum_k \dot{p}_{s_1}(\alpha_k+\Lambda)+
\dot{p}_{s_2}(\alpha_k-\Lambda)}
(this corresponds to antiferromagnetic interactions. For the ferromagnetic case
switch the sign of $E$). The momentum of the corresponding state reads
\eqn\mom{P_I=\sum_k p_{s}(\alpha_k+\Lambda)+p_{s}(\alpha_k-\Lambda)}
and
\eqn\momI{P_{II}=\sum_k p_{s_1}(\alpha_k+\Lambda)+p_{s_2}(\alpha_k-\Lambda)}

\subsec{Physical comments}

Consider first array  I.  Call $\Delta$ the spatial period of the system (equal
to
two lattice spacings in the original square lattice). Call $\tau$ the
translation operator $j\rightarrow j+1$ so $\tau^2={\cal T}$. Then
$t^s(i\Lambda/2)
\tau$ and $\tau(t^s(-i\Lambda/2))^{-1}$ describe vertical propagation as in the
following
figure
\bigskip
\noindent
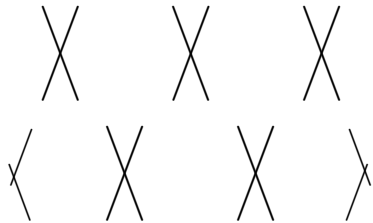
\centerline{\hbox{\rlap{\raise30pt\hbox{$\hskip-2cm\Biggl/\hskip-.45cm\Biggr\backslash
\hskip1cm\Biggl/\hskip-.45cm\Biggr\backslash\hskip1cm\Biggl/\hskip-.45cm
\Biggr\backslash$}}
\rlap{\lower8pt\hbox{$\hskip-1.4cm\Biggl/\hskip-.45cm\Biggr\backslash
\hskip1cm\Biggl/\hskip-.45cm\Biggr\backslash$}}
\rlap{\lower14pt\hbox{$\hskip-2.6cm\Bigr\backslash$}}
\rlap{\lower3pt\hbox{$\hskip-2.7cm\Bigl/$}}
\rlap{\lower14pt\hbox{$\hskip.9cm\Bigl/$}}
\rlap{\lower3pt\hbox{$\hskip.82cm\Bigr\backslash$}}}}
\bigskip
\noindent

\noindent while $t^s(i\Lambda/2)$ and $(t^s(-i\Lambda/2))^{-1}$ describe
propagation  along the right and left diagonal. The system  can thus be
intuitively
considered
as a "lattice light-cone" where  the vertices represent
 interactions  between $s+1$  "bare" particles (see \ref\DV{C.Destri,
H. de Vega, J.Phys. A22 (1989) 1329.} for a related approach). The natural
hamiltonian
associated with this description is given by $H_I={1\over
i}\ln\left[t^s(i\Lambda/2)
(t^s(-i\Lambda/2))^{-1}\right]$. It is however non local, and this causes
some difficulties (see later).
Array II would have a similar interpretation
 but with the left (right)
diagonal carrying spin $s_1$ ($s_2$). Note that in this case the  only
interactions
 are between left and right moving bare particles.
Let us emphasize
 that in such a picture
we are dealing with quantum mechanics at real time.

In the limit
$\Lambda\rightarrow\infty$ the first term in \hai\ and \haii\ decays
exponentially while
the second one remains of order one. In that limit, the second term in \hai\
and \haii\
describes a RSOS hamiltonian with some special boundary conditions, coupling
respectively only even (odd) sites. In the limit
$\Lambda\rightarrow\infty$ the hamiltonian \hamiii\ is thus the sum of two
decoupled
RSOS hamiltonians. When $\Lambda$ is finite, they interact due to the first
terms
in \hai\ and \haii\ . It is easy to see on the other hand that the transfer
matrix
$t^s(i\Lambda/2)(t^s(-i\Lambda/2))^{-1}$ becomes trivial in the limit
$\Lambda\rightarrow\infty$.

\newsec{Thermodynamic limit}

We now consider the thermodynamic limit $N\rightarrow\infty$. For the
computation of thermodynamic quantitites,
 it is expected that the only relevant solutions of \ba\ are given in
terms of strings \ref\TS{M.Takahashi, M.Suzuki, Prog. Theor. Phys. 48 (1972)
2187.}. For technical simplicity we will restrict in the following to values
 of the anisotropy parameter $q=\exp(i\gamma)$, where $\gamma=\pi/t$, $t$
 being an integer.

\subsec{XXZ type quantum chains}

For the vertex or generalized XXZ model case  the solutions of \ba\
are made of $n$ strings of
length $n=1,\ldots,t-1$ and of the ($n=1^-$) antistring of length 1 \TS\
(recall
that an $n$ string is a set of $n$ complex numbers
$\alpha_k^{(l)}=\alpha_k^n+il\gamma$
with $l=n-1,n-3,\ldots,-n+1$ and an antistring of length one is a complex
number with imaginary part equal to $\pi$).
 Regrouping the numbers $\alpha$ into their various strings leads to
equations similar to \ba\ but for the (real) centers of the strings
\eqn\baI{\prod_{j=1}^{j=N}\exp\left[ip_{n, s_j}(\alpha_k^n-2iu_j)\right]
\prod_{ml}\exp\left[i\phi_{n,m}(\alpha^n_k-\alpha^m_l)\right]=-1,}
where the second product is taken over all possible strings and then over
 each center. In the thermodynamic limit equation \baI\ gives the
following integral equations for  the densities $\rho_n,
\tilde{\rho}_n$ of strings and  holes
\eqn\baII{\sum_{s_j}\dot{p}_{n, s_j}\star {\cal P}_{s_j}+\sum_m\dot{\phi}_{n,m}
\star\rho_m=2\pi\epsilon(\rho_n+\tilde{\rho}_n).}
where ${\cal P}_s$ encodes the distribution of the heterogeneous spectral
 parameters for the spins $s$ (eg a sum of two delta functions  or  a single
delta function in  the array I or II) and $\epsilon=-1$ for the antistring.

In the following we set $a_{n,s}\equiv\dot{p}_{n,s}$
 and $A_{n,m}\equiv2\pi\delta_{n,m}\delta-\dot{\phi}_{n,m}$.
The various above functions are more conveniently expressed in terms of
 the Fourier transforms of their derivatives. For $n,s=1,\ldots,t-1$ one has
\eqn\for{\hat{a}_{n,s}={\sinh( nx)\sinh[(t-s)x]\over\sinh (tx)\sinh(x)},
\ n\leq s.}
and
\eqn\forI{\hat{A}_{n,m}=2{\cosh(x)\sinh(nx)\sinh[(t-m)x]\over\sinh(tx )
\sinh(x)},\ n\leq m.}
when one or two of the labels involve the antiparticle one has
\eqn\forII{\hat{a}_{1^-,n}=-{\sinh (sx)\over \sinh (tx)},}
and
\eqn\forIII{\hat{A}_{1^-,n}=-2\cosh(x){\sinh(nx)\over \sinh(tx)},\ n\neq 1^-,\
 n\neq t-1,}
\eqn\forIIII{\hat{A}_{1^-,t-1}=-{\sinh[(t-2)x]\over \sinh(tx)},\
\hat{A}_{1^-,1^-}=\hat{A}_{11}}
and the kernels are symmmetric otherwise.

Once the thermodynamic limit has been taken we compute the density of
 free energy of
 the infinite chain (with still finite lattice spacing)
 ${\cal F}={\cal E}-T{\cal S}$, where $T$ is the  temperature and ${\cal E},
{\cal S}$ are the densities of  energy and entropy respectively.
The formula for the entropy is the $N\rightarrow\infty$ limit of the
combinatorial
entropy of distribution of holes and particles \ref\YY{C.N.Yang, C.P.Yang,
 Phys. Rev. 147 (1966) 303.}
\eqn\ther{{\cal S}=\sum_{n}\int_{-\infty}^{\infty}\left[(\rho_n+
\tilde{\rho}_n)\log(\rho_n+\tilde{\rho}_n)-\rho_n\log(\rho_n)-\tilde{\rho}_n
\log(\tilde{\rho}_n)\right].}
The expression for the energy depends on the particular array. For array I one
has

\eqn\therI{{\cal E}_I=-{2\over t}\sum_n\int_{-\infty}^{\infty}
\left[a_{n,s}(\alpha+\Lambda)+a_{n,s}(\alpha-\Lambda)\right]\rho_n(\alpha)d\alpha,}
and for  array II
\eqn\therI{{\cal E}_{II}=
-{2\over t}\sum_n\int_{-\infty}^{\infty}\left[a_{n,s_1}(\alpha+\Lambda)+
a_{n,s_2}(\alpha-\Lambda)\right]\rho_n(\alpha)d\alpha,}

\subsec{RSOS quantum chains}

We can follow the same steps for the RSOS chain. The additional  phases
in \barsos\  seem to disappear in the equations for densities in the
thermodynamic
limit, but in fact as discussed in \BR\ , they suppress the $t-1$ string and
the antistring. This implies moreover that for the $t-2$ string there are no
 holes. One can then eliminate $\rho_{t-2}$ in the various equations to get
 a system exactly identical to \baII\ but where $n=1,\ldots,t-3$ and where
$t$ is replaced by $t-2$ in the various kernels
\BR\ .

\subsec{Thermodynamic equilibrium  for XXZ type quantum chains}

The minimization of the free energy leads to coupled equations
 for the densities  \TS\ \YY\ .
For   array I we have the following.
 Introducing $\epsilon_n=T\log(\tilde{\rho}_n/\rho_n)$ we get, after inversion
 of the matrix in \baII ,
\eqn\tba{-e(\alpha)\delta_{n,s}=\epsilon_n-Ts\star\sum_m N^{XXZ}_{n,m}
\log[1+\exp(\epsilon_m)/T].}
where we have defined
\eqn\dI{e(\alpha)\equiv s(\alpha+\Lambda)+s(\alpha-\Lambda).}
and
\eqn\dII{s(\alpha)\equiv{1\over \cosh(t\alpha/2)}.}
In the last expression $N^{XXZ}_{n,m}$ is the adjacency matrix of the following
 diagram

\bigskip
\noindent
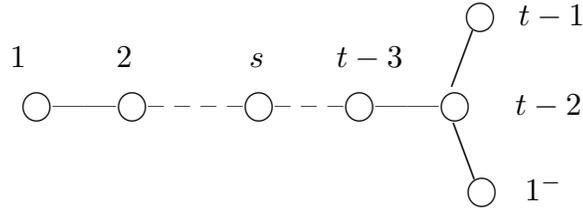
\centerline{\hbox{\rlap{\raise28pt\hbox{$\hskip5.5cm\bigcirc\hskip.25cm t-1$}}
\rlap{\lower27pt\hbox{$\hskip5.4cm\bigcirc\hskip.3cm 1^-$}}
\rlap{\raise15pt\hbox{$\hskip5.1cm\Big/$}}
\rlap{\lower14pt\hbox{$\hskip5.0cm\Big\backslash$}}
\rlap{\raise15pt\hbox{$1\hskip1cm 2\hskip1.3cm s\hskip.8cm t-3$}}
$\bigcirc$------$\bigcirc$-- -- --
--$\bigcirc$-- -- --$\bigcirc$------$\bigcirc$\hskip.5cm $t-2$ }}

\bigskip
\noindent

\noindent For the free energy per unit length we find
\eqn\fren{F_I=-\int_{-\infty}^{\infty}\left\{T\log[1+\exp(\epsilon_s/T)]
+a_{s,s}(\alpha)\right\}e(\alpha)d\alpha.}

Similarly for the  array II, inversion of the matrix in \baII\ now leads
 to
\eqn\ms{-e_{s_1}(\alpha)\delta_{n,s_1 }-e_{s_2}(\alpha)\delta_{n,s_2} =
\epsilon_n-Ts\star\sum_m  N^{XXZ}_{n,m}\log[1+\exp(\epsilon_m/T)],}
where
\eqn\ede{e_{s_1}(\alpha)=s(\alpha+\Lambda)
,\ e_{s_2}(\alpha)=s(\alpha-\Lambda),}
and for the free energy
\eqn\frenI{F_{II}=-\int\{T\log[1+\exp(\epsilon_{s_1}/T)]+a_{s_1s_1}\}
e_{s_1}(\alpha)d\alpha+\{T\log[1+\exp(\epsilon_{s_2}/T)]+a_{s_2s_2}\}
e_{s_2}(\alpha)d\alpha.}

\subsec{Thermodynamic equilibrium for RSOS type quantum chains}

 Things are very similar in that case. The only difference is that
sums over solutions run only over $n=1,\ldots,t-3$ strings in \tba\ , \ms\
, \fren\ and \frenI\ while  in \tba\ and \ms\ $N^{RSOS}_{n,m}$ is the
incidence matrix of the
diagram

\bigskip
\noindent
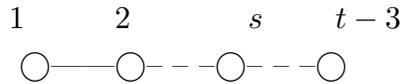
\centerline{
\rlap{\raise15pt\hbox{$1\hskip1cm 2\hskip1.3cm s\hskip.8cm t-3$}}
$\bigcirc$------$\bigcirc$-- -- --$\bigcirc$-- -- --$\bigcirc$ }

\bigskip
\noindent

\smallskip
\noindent {\sl 3.5 $T\rightarrow 0$ limit for array I}

 Consider first the   ground state of array I at $T=0$ (in this paragraph
the distinction between generalized XXZ and RSOS chains is simply encoded in
 the two possible incidence diagrams).
 System \tba\ reads then
\eqn\exct{-e(\alpha)\delta_{n,s}=\epsilon_n-s\star\sum_m N_{n,m}\epsilon_m^+,}
where $\epsilon^+(x)=\epsilon(x)H(\epsilon(x))$ is the positive part of the
 $\epsilon$ function (H being the Heaviside step function).
The solution of this system is $\epsilon_n^+=0$ and
$\epsilon_n^-=-e\delta_{ns}$. From \exct\ we see that the ground state
is filled with $s$ strings. Excitations above this ground state are made
of holes in the sea of $s$ strings, with excitation energy precisely
equal to $e$ (other strings are important in determining for instance the
spin of the excitations, but they do not contribute to the energy and
momentum).
 The momentum of the excitations is similarly found to be
\eqn\pp{p=2\tan^{-1}\left({\sinh t\alpha/2\over\cosh t\Lambda/2}\right)+\pi.}
and as in \ref\FT{L.D.Faddeev, L.A.Takhtajan, Phys.
 Lett. 85A (1981) 375.} they always occur in pairs (with our convention
$p$ is defined modulo $4\pi$). There are two Brillouin zones. In the first
 one the model has
massive excitations with a gap at $p=\pi$ that vanishes exponentially
 with $\Lambda$.
In the second zone the model has massless excitations  around
$p=0,2\pi$.
 This is true both for XXZ and
RSOS chains provided $s\leq t-3$. In our subsequent discussion we discard the
finite parts of the momenta. The role of these contributions in the scaling
limit is similar to the one in spin $s=1$ case discussed eg in
\ref\Aff{I.Affleck,
 in "Fields, Strings and Critical Phenomena", Les Houches Lectures,
E.Brezin et al. Editors, North Holland (1986).}

\subsec{$T\rightarrow 0$ limit for array II}

 In that case the ground state is  filled with $s_1$ and $s_2$ strings and the
 excitations above it are made of holes in the sea, with excitation energies
 equal
to $e_{s_1},e_{s_2}$. The momentum of excitations is similarly found to be
\eqn\ppii{p_{s_1}=2\tan^{-1}\{\exp[t(\alpha+\Lambda)/2]\},\
p_{s_2}=2\tan^{-1}\{\exp[t(\alpha-\Lambda)/2]\}.}
There are massless excitations around $p=0,\pi$ in both cases.
Again this is true
for both XXZ and RSOS chains provided $s_1,s_2\leq t-3$.

\newsec{The  scaling limit}

\subsec{XXZ type chain and array I}

In the first Brillouin zone the dispersion relation exhibits a
 gap that vanishes in the
limit
  $\Lambda\rightarrow\infty$ as
\eqn\eg{m\Delta\equiv\delta E\approx 4\exp(-t\Lambda/2),}
Let us  consider the  scaling limit. Let $N>>1$
 and at the same time $\Delta<<1$ in such a way that the dimensional
 length of the chain $L=\Delta N$ is much bigger than the correlation length
 $\xi\equiv{\Delta\over\delta E}$, the latter itself remaining finite. Hence
 we need
\eqn\sccon{\Lambda,N>>1,\ \Delta<<1,\ N\delta E>>1,\
\Delta\propto\delta E}
We then concentrate on the contributions of order $\Delta$ to the hamiltonian.
At finite rapidity, that is very close to the origin of the first Brillouin
zone,
 we then find for the excitation energies
\eqn\enlim{e(\alpha)\approx 4e^{-t\Lambda/2}\cosh(t\alpha/2)\equiv \Delta
 m\cosh\theta,}
where the mass $m\equiv 1/\xi$, and the physical rapidity $\theta\equiv
t\alpha/2$. Similarly for the momentum
\eqn\pplim{p(\alpha)\approx 4e^{-t\Lambda/2}\sinh(t\alpha/2)\equiv \Delta
 m\sinh\theta,}
Relations \enlim\ and \pplim\ correspond to  relativistic particles of mass
$m$.
Excitations at finite distance from the origin in the first Brillouin zone
 have finite
energy and will disappear in the scaling limit.

In the second Brillouin zone we also have excitations with energy of
order $\Delta$.
As $\alpha>>\Lambda$ ($\Lambda$ being itself very large) we have
\eqn\addi{e_L(\alpha)\approx 2e^{t\Lambda/2}e^{-t\alpha/2},}
with
\eqn\addii{p_L(\alpha)\approx -e_L.}
Similarly as  $\alpha<<-\Lambda$ we have
\eqn\addiii{e_R(\alpha)\approx 2e^{t\Lambda/2}e^{t\alpha/2},}
with
\eqn\addii{p_R(\alpha)\approx e_R.}
As in \FT\ there are only even number of the different types of
excitations and the finite contributions to the momenta can be discarded.
These relation
correspond to massless left (respectively right) moving particles.

As $\Delta\rightarrow 0$ we expect  the following
\eqn\hla{H_{latt}\approx N{\cal E}_0+H_{scal}\Delta,}
where $H_{scal}$ is the hamiltonian of the continuum field theory obtained in
the scaling limit. On the other hand from \ref\ZamoII{Al.B.Zamolodchikov,
 Phys. Lett. B253 (1991) 391.}, \ref\FS{P.Fendley, H.Saleur, Nucl. Phys. B388
 (1992) 609.} we observe that
the thermodynamic equations \tba\   of our   system, for energies of order
$\Delta$,
  coincide  in
the first Brillouin zone with the so called Thermodynamic Bethe Ansatz
 equations\foot{Here by TBA we refer to thermodynamic analysis of a gas
of relativistic particles in one dimension, with factorized scattering.}
for
 the spin $s$ or fractional supersymmetric sine-Gordon model
 \BL\ with coupling constant
\eqn\bet{{\beta^2_{SG}\over 8\pi}={\bar {t}\over s(s+\bar{t})},}
where $\bar{t}=t-s$ and $c_{UV}={3s\over s+2}$, once  the temperature
in our problem has been identified as usual with the inverse  radius
 $R$ in \ZamoII\ , \FS\ . Similarly in the second Brillouin zone
the thermodynamic equations  of our system for energies of order $\Delta$
coincide with the "massless TBA" equations for the generalized Gaussian model
(a system
made of a free boson and $Z_s$ parafermions) at radius of
compactification appropriately related to \bet\ \ref\FSI{P.Fendley, H.Saleur,
"Lectures on massless flows", preprint USC -93-022.}. The excitations are then
massless since one is describing a conformal field theory. Moreover, as
$\Lambda>>1$,
the regions of interest in the first and second Brillouin zone completely
decouple
since the kernel in the lattice equation \tba\ decays exponentially outside a
region of finite size in $\alpha$ variable.
We therefore conclude that
\eqn\addi{H_{scal}=H_{SG-s}+H_{G-s},}
where the first term is the hamiltonian of the spin-s sine-Gordon model
and the second term the hamiltonian of the generalized Gaussian model (a
conformal
invariant
 theory) on a cylinder of circumference $L$ with appropriate boundary
conditions: $H_{G}={2\pi\over L}(L_0+\bar{L}_0-{1\over 12})$.

  The identification of the scaling limit could  be made
 completely by computing, from the lattice model \KR\ \ref\Ko{V.Korepin,
Th. Mat. Phys. 41 (1979) 169.}\ref\TW{A.Tsvelick, P.Wiegmann, Adv.
 in Physics 32 (1983) 453.},  the $S$ matrix for the scattering of
 excitations. For doing so one interprets the holes in the sea as
the physical particles.
 The other string solutions correspond then to pseudoparticles. In the present
case three S matrices would have to be computed: the one describing the
massive SG model and two others, identical ones, describing
scattering between left (respectively right) "massless particles" in the
conformal field theory. One would thus get three TBA diagrams like

\bigskip
\noindent
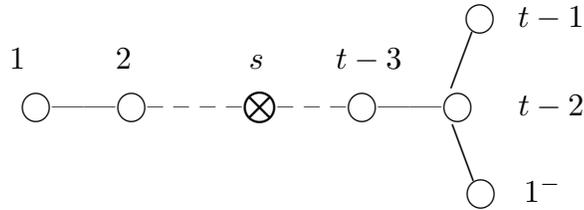
\centerline{\hbox{\rlap{\raise28pt\hbox{$\hskip5.5cm\bigcirc\hskip.25cm t-1$}}
\rlap{\lower27pt\hbox{$\hskip5.4cm\bigcirc\hskip.3cm 1^-$}}
\rlap{\raise15pt\hbox{$\hskip5.1cm\Big/$}}
\rlap{\lower14pt\hbox{$\hskip5.0cm\Big\backslash$}}
\rlap{\raise15pt\hbox{$1\hskip1cm 2\hskip1.3cm s\hskip.8cm t-3$}}
$\bigcirc$------$\bigcirc$-- -- --
--$\bigotimes$-- -- --$\bigcirc$------$\bigcirc$\hskip.5cm $t-2$ }}

\bigskip
\noindent where the marked node  corresponds to a massive particle of mass $m$
, repsectively to a left or
right moving particle.  We comment more on massless scattering in section 5
where we also  give a detailed  example of an S matrix  computation. Let us
  just mention that the $S$ matrices   extracted  in this fashion
coincide with
 the known one  for spin $s$ sine-gordon \BL\  and the ones for
generalized Gaussian models \FSI\ \FT\ \ref\ZamoZamo{A.B.Zamolodchikov,
Al.B.Zamolodchikov, "Massless factorized scattering and sigma models with
topological terms, preprint ENS-LPS-335-91.}.

To understand  more directly what perturbed conformal field theory we are
 dealing with we finally consider  \ref\BR{V.V.Bazhanov,
 N.Yu Reshetikhin, Prog. Th. Phys. 102 (1990) 301.}
a  limiting process similar to \sccon\
 but with the correlation length much bigger than $L$ ie
\eqn\scconI{\Lambda,N\rightarrow \infty,\ \Delta\rightarrow 0,\ N\delta E<<1,\
\Delta\propto\delta E}
In that limit we have
\eqn\pert{H_{scal}=H_{conf}+\exp(-\Lambda)H_{pert},}
where $H_{conf}=H_{G-s}+H_{G-s}$ (this corresponds to \hamiii\ being a sum of
two decoupled XXZ type models in the limit $\Lambda\rightarrow\infty$)
and $H_{pert}$ is determined by the first term in \hai\ and \haii\ . Comparing
this
with the formula giving the physical mass leads to the value of the conformal
 weight of the perturbation $h_{SG-s}={t-1\over t}$.
This conformal weight could also be found for $s=1$
 by noticing that  taking the limit
 $\Lambda\rightarrow\infty$ the equations \ba\  with $s=1$ reduce
to the ones of the Thirring model with bare mass
$m_0=4\sin\gamma\exp(-\Lambda)$ and bare rapidity
$\alpha$ \Ko\ \ref\BT{H.Bergknoff, H.B.Thacker, Phys. Rev. D19 (1979) 3666.}.
Notice that the conformal weight does not depend on the spin $s$, ie it
 depends only on the skeleton of the TBA diagram and not on which node(s)
 are massive. This result can also be directly established at the level of
 the TBA equations \ZamoI .

\subsec{RSOS  chain and array I}

Things are quite similar in that case, with the additional
truncation of \tba\ . The thermodynamic equations
\tba\ for energies of order $\Delta$ now coincide in the first Brillouin
zone with the TBA equations for the theory ${\cal MA}_{st}^-$ written in \Zamo\
{}.
Similarly in the second Brillouin zone the equations coincide with  the
massless
TBA that describes minimal conformal field theories \FSI\ . These three TBA are
associated with the same diagram

\bigskip
\noindent
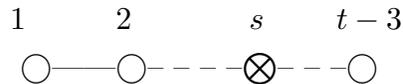
\centerline{
\rlap{\raise15pt\hbox{$1\hskip1cm 2\hskip1.3cm s\hskip.8cm t-3$}}
$\bigcirc$------$\bigcirc$-- -- --
--$\bigotimes$-- -- --$\bigcirc$ }

\bigskip
\noindent where the marked node corresponds to a massive particle of mass $m$,
respectively a left or right moving particle.
Equation   \addi\ now reads
\eqn\adI{H_{scal}=H_{{\cal MA}_{st}^-}+H_{{\cal M}_{st}},}
where  the notations are explained in the introduction. In the limit \scconI\
one finds
\eqn\pertI{H_{scal}=H_{conf}+\exp(-2\Lambda)H_{pert},}
where $H_{conf}=H_{{\cal M}_{st}}+H_{{\cal M}_{st}}$, $H_{pert}$ is determined
by the first term in \hai\ \haii\ . The main difference with the
unrestricted case discussed above is that the perturbation
starts with order $\exp(-2\Lambda)$ instead of order $\exp(-\Lambda)$: this
is due to the different spectral parameter dependence of the R-matrices
(the different  gradations from $\hat{U}_qsl(2)$ point of view
\ref\FL{G.Felder, A.Leclair, in "Infinite Analysis", A.Tsuchiya et al.
 Eds., World Scientific, Singapore (1991).})\foot{The difference
 of exponent in \pert\ and \pertI\ can
also be observed in the ellpitic case: see the Boltzmann weights
 in \ABF\ .}. As a consequence  one finds $h={t-2\over t}$
 as announced in the introduction.

To conclude this subsection  we have  established that array I provides
 a regularization of the interpolating theory ${\cal M}A_{s,t}^-$ (plus a
"background" ${\cal M}_{st}$ theory that decouples).

\subsec{XXZ type chain and array II}

We consider again the scaling limit defined in \sccon\ . In the region
$|\alpha|
>>\Lambda$ (the second Brillouin zone) things are the same as for array I.
They are quite different in the first Brillouin zone. In the present case
we have, in the limit $\Lambda\rightarrow\infty$, and for finite rapidities,
\eqn\enms{e_{s_1}\approx {\Delta m\over 2}e^{-\theta},\ e_{s_2}\approx
{\Delta m\over 2}e^
{\theta},}
where $m,\theta$ are the same as before. It is straightforward to study as well
 the momentum of the excitations and one finds
\eqn\pms{p_{s_1}=-e_{s_1},\ p_{s_2}=e_{s_2},}
ie the dispersion relations for left and right moving massless particles. We
thus
have left and right moving massless particles in both Brillouin zones. However,
while
in the second zone there is no left-right scattering since the two types
of particles are widely separated in the rapidity variable ($\alpha>>\Lambda$
and $\alpha<<-\Lambda$), in the first zone
we expect to have non trivial left right scattering. This leads to a theory
that is non scale invariant.
In general however,  one does not get a meaningful scaling limit due to the
lack of
 symmetry between left and right sectors.  The first step to recover this
symmetry is to consider the RSOS version.

\subsec{RSOS chain and array II}

In the scaling limit, thermodynamic equations
in the first Brillouin zone coincide with then with the TBA  equations
associated with a diagram
 of type $A_{n-3}$ with physical particles at nodes $s_1$ and $s_2$.  For LR
symmetry one
then has to require $s_1\equiv s=t-2-s_2$, leading to the TBA diagram
\bigskip
\medskip\noindent
\centerline{\hbox{\rlap{\raise15pt\hbox{$\hskip.2cm 1\ldots \hskip.47cm
s\ldots
\hskip.3cm t-2-s\ldots \hskip.2cm t-3$}}
\rlap{\lower15pt\hbox{$\hskip1.2cm {m\over 2}e^{-\theta}\hskip.7cm{m\over 2}e^
\theta$}}
$\bigcirc$-- -- --$\bigotimes$-- -- -- --$\bigotimes$-- -- --$\bigcirc$}}
\bigskip
\noindent

This TBA coincides then with  the one conjectured in \Zamo\ ,
\ref\ZamoIII{Al.B.Zamolodchikov, Nucl. Phys. B358 (1991) 497.} for the flow
 between successive minimal models. Hence we expect
\eqn\thi{H_{scal}=H_{{\cal MA}_{st}^+}+H_{{\cal M}_{st}}.}

 Observe  on the other hand  the following important symmetry for a RSOS
model with $t-1$ heights (with appropriate normalization)
\eqn\rsym{R_{RSOS}^{s,s}(u)=R_{RSOS}^{s,t-2-s}\left[u+{\pi\over 2}\right].}
(this can be derived from identities in \KR\ together with the observation that
 in the latter reference the spectral parameter $u$ in the weights $W^{pq}$
is shifted, when compared to ours, by the amount ${p-q\over 2}\gamma$).

 Hence
 array II in the RSOS case is equivalent to  array I, but with the cutoff
$\Lambda$
acquiring an imaginary part $i{\pi\over 2}$.
We can now  make the identification with perturbed minimal models more complete
 by considering  the limit \scconI\ . Due to the imaginary part of $\Lambda$ we
observe indeed the change of sign of the  coupling constant in \pertI\ for the
 RSOS  model:
\eqn\thii{H_{scal}=H_{{\cal M}_{st}}+H_{{\cal
M}_{st}}-\exp(-2\Lambda)H_{pert},}
with $H_{pert}$ the same as in \pert\ .
These results indicate that we have a regularization of the interpolating
theory ${\cal M}A^+_{s,t}$. To complete the identification we shall
compute the $S$ matrix in the following section.

 Notice that, based on \rsym\ , it is tempting to consider also a type I array
 for the vertex model with a cutoff $\Lambda$ having imaginary part
${i\pi\over 2}$. From \pert\ it corresponds indeed to the  spin $s$
sine-Gordon model with imaginary coupling constant. More detailed study of
 this model  should be of interest \ref\SS{S.Skorik, H.Saleur, in
preparation.}.

\subsec{Computation of central charges}

In the framework of the TBA it is standard to compute the central charge
in the UV and IR. Let us notice that it is also possible to
extract these two numbers directly from the lattice model by considering
appropriate limits. In the scaling limit we expect the entropy ${\cal
S}(T,\Lambda)$
to become a function of the scaling variable ${m\over \tau}$ where
$\tau={T\over\Delta}$. The UV limit corresponds to $\tau>>m$ and the IR
limit to $\tau<<m$. To extract the central charge from the lattice model
one has to proceed as follows
\eqn\sihep{{\cal S}(T,\Lambda)\approx {\pi T\over 3}c_{IR}\hbox
{ if }T<<e^{-t\Lambda/2},
\ T\rightarrow 0,\ \Lambda\rightarrow\infty,}
and
\eqn\siihep{{\cal S}(T,\Lambda)\approx {\pi T\over 3}c_{UV}\hbox{ if
}T>>e^{-t\Lambda/2},
\ T\rightarrow 0,\ \Lambda\rightarrow\infty,}
For the RSOS model and array I the results of this computation are
$c_{UV}=2\left(
1-{6\over t(t-1)}\right)$ and $c_{IR}=1-{6\over t(t-1)}$ while  for array II
$c_{UV}$ is the
same but $c_{IR}=1-{6\over t(t-1)}+1-{6\over (t-1)(t-2)}$.

\newsec{The $S$ matrix for massless flows between minimal models}

\subsec{Massless S matrices}

To start we briefly discuss the definition of massles S matrices. For
integrable systems
one can define scattering amplitudes via symmetry properties of eigenstates of
the
hamiltonian of the system under permutation of particles. We parametrize energy
and
momentum of left and right moving massless particles as in \enms\ and \pms\ so
if
$\Psi_{\theta^R,\theta^L}$ is an eigenstate
of $H$ and $P$ one has
\eqn\kkki{H\Psi_{\theta^R,\theta^L}={m\over 2}\left(
\sum_ie^{\theta_i^R}+\sum_je^{-\theta_j^L}\right)\Psi_{\theta^R,\theta^L},}
and
\eqn\kkkii{P\Psi_{\theta^R,\theta^L}={m\over 2}\left(
\sum_ie^{\theta_i^R}-\sum_je^{-\theta_j^L}\right)\Psi_{\theta^R,\theta^L},}
Consider a given set of rapidities $\theta_1^R,\ldots,\theta_N^R,\theta_1^L,
\ldots,\theta_M^L$ and order them
$\theta_{\sigma(1)}\leq\ldots\leq\theta_{\sigma(N+M)}$ where $\sigma\in
S_{N+M}$.
To a given  order characterized by some element $\tau$ of $S_{N+M}$
one can associate a  linear space ${\cal H}_\tau(\theta^R,\theta^L)$ of
dimension $C^{D(N+M)}$
where $D$ is the number of isotopic degrees of freedom. All these spaces
are isomorphic and  the
isomorphism
\eqn\iso{{\cal H}_\tau(\theta^R,\theta^L)\approx
{\cal H}_{\tau'}(\theta^R,\theta^L),\ \tau,\tau'\ \in\ S_{N+M}}
defines some S matrix $S_{\tau,\tau'}(\theta^R,\theta^L)$. The two
particle S matrix is obtained in the case $\tau=\sigma$ and
$\tau'=\sigma\sigma_i$
where $\sigma_i$ transposes $i$ and $i+1$ as
\eqn\isoi{S_{\sigma,\sigma_i\sigma}=S_{RR}(\theta_j^R-\theta_k^R)\hbox{ if }
\theta_{\sigma(i)}=\theta^R_j,\ \theta_{\sigma(i+1)}=\theta^R_k,}
$$
S_{\sigma,\sigma_i\sigma}=S_{LL}(\theta_j^L-\theta_k^L)\hbox{ if }
\theta_{\sigma(i)}=\theta^L_j,\ \theta_{\sigma(i+1)}=\theta^L_k,
$$
and
$$
S_{\sigma,\sigma_i\sigma}=S_{LR}(\theta_j^L-\theta_k^R)\hbox{ if }
\theta_{\sigma(i)}=\theta^L_j,\ \theta_{\sigma(i+1)}=\theta^R_k,
$$
with a similar equation for $S_{RL}$.

When we put the theory on a line of length $L$ this symmetry
implies the form of the eigenfunctions
\eqn\isoii{\Psi_{\theta^R,\theta^L}(x_1,\ldots,x_{N+M})=\sum_{\tau\in S_{N+M}}
A_\tau\exp[i(p_{\sigma(1)}x_1+\ldots+p_{\sigma(N+M)}x_{N+M})],}
where the amplitudes $A_{\tau}$ behave as \iso\ . The standard form of periodic
boundary conditions \FSI\ then follows.

\subsec{S matrix for ${\cal MA}_{st}^+$}

To get the $S$ matrix of the excitations we proceed as in \KR\ \TW\ .
 For simplicity we first consider the case $s=1$. Start from
\eqn\sI{{1\over 2}\left[a^{(t-2)}_{n}(\alpha+\Lambda) +a^{(t-2)}_{n,t-3}
(\alpha-
\Lambda)\right] =2\pi \tilde{\rho}_n+\sum_{m=1}^{t-3}A^{(t-2)}_{n,m}\star\rho_m
,}
where the subscript mean $t$ is replaced by $t-2$ in equations
\for\ up to \forIIII\ . Rename now $\sigma_L\equiv \tilde{\rho}_1,\
\sigma_R\equiv\tilde{\rho}_{t-3}$.
 The densities $\sigma_{L,R}$ now are considered as describing physical
 particles (they are  densities of holes  in the original Bethe ansatz due to
 the structure of the ground state and its excitations). Similarly rename
 $\sigma_n\equiv\rho_{n-1},\ e_L\equiv e_1,\
e_R\equiv e_{t-3}$ and isolate the physical densities. After lengthy
calculation one finds
\eqn\sII{2\pi
\tilde{\sigma}_n+\sum_{m=1}^{m=t-5}A^{(t-4)}_{m,n}\star\sigma_m=a^
{(t-4)}_n\star\sigma_L+a^{(t-4)}_{n,t-5}\star\sigma_R}
as well as
\eqn\sIII{2\pi (\sigma_L+\tilde{\sigma}_L)=e_L+a^{(t-4)}_1\star\sigma_L+
a^{(t-4)}_{t-5}\star\sigma_R-\sum_{n=1}^{n=t-5}a^{(t-4)}_n\star\sigma_n,}
and
\eqn\sIIII{2\pi (\sigma_R+\tilde{\sigma}_R)=e_R+a^{(t-4)}_{t-5}\star\sigma_L+
a^{(t-4)}_1\star\sigma_R-\sum_{n=1}^{n=t-5}a^{(t-4)}_{t-5,n}\star\sigma_n,}
here $a,A$ refer to the same kernels as in \forI\ \forIIII\ but with $t$
 replaced by $t-4$. As a consequence there is $L,R$ symmetry ie
$A^{(t-4)}_{n,m}=A^{(t-4)}_{t-2-n,t-2-m}$. Taking now the limit
 $\Lambda\rightarrow\infty$, the equations \sII\ \sIII\ \sIIII\  can be
identified with the ones one would write for a theory with L particles of
 spin 1, R particles of spin $t-3$ with factorized scattering given
 by
$S_{LL}=S^{11}_{t-3}, S_{RR}=S^{t-5,t-5}_{t-3}=S_{LL}$ and
 $S_{LR}=S^{1,t-5}_{t-3}$
\foot{These identifications hold up to a phase, which cannot be
 determined from the TBA equations}, where
in each case $S$ is made of Boltzmann weights of the appropriate RSOS model
 with $t-3$ heights\foot{To be put in correspondence with the minima in
the Landau Ginzburg picture} and adjacency conditions depending on the spin,
 and the usual minimal prefactors ensuring unitarity and crossing symmetry
\foot{These identifications hold up to a phase, which cannot be
 determined from the TBA equations}. Indeed \sII\
just expresses the densities of pseudoparticles in the problem of computing
 the eigenvalues for passing a L or R particle through a set of
 L and R particles with the densities $\sigma_L,\sigma_R$ (that is the
 eigenvalues of the "monodromy matrix" $T^s$ in section 2). Similarly \sIII\
and \sIIII\
express
 the periodicity condition for the wave function of L and R particles.

Using the symmetry between spin 1 and spin $t-5$ in the RSOS version,
one can rewrite $S_{LR}(\theta)$ as $S^{11}\left[\theta+i
{(t-2)\pi\over 2}\right]$ as well, in agreement with the result conjectured
 in \Fetc.

The computation easily extends to the higher $s$ case. One finds
then
\eqn\mri{S_{LL}=S^{11}_{s+1}\otimes S^{11}_{t-1-2s}\otimes 1,\ S_{RR}=1\otimes
S^{11}_{t-1-2s}\otimes S^{11}_{s+1},}
and
\eqn\mrii{S_{LR}=1\otimes S^{11}_{t-1-2s}\left[\theta+i
{(t-2)\pi\over 2}\right]\otimes 1,}
 where $S_{n}$ is the S matrix based on the
RSOS model with $n$ heights. In the above notation, left and right components
of the
tensor product correspond respectively to L and R RSOS models
(with $s+1$ heights)
, while the middle
component is as in the case $s=1$ above, ie it contains both  L
and R RSOS models (with $t-1-2s$ heights),
and the S matrix acts on LL, RR, LR respectively.

\subsec{Physical comments}

 Notice that in the interacting
 theory we have  non trivial LL,RR,LR scattering, while in the
lattice light-cone picture  only LR scattering occurs between the bare
 particles.

It may be useful to summarize the various correspondence between
lattice RSOS models and thermodynamic equations, RSOS models in
 S matrices, and TBA. In all the previous sections
the lattice model
is based on quantum parameter $q=\exp(i\pi/t)$.  This corresponds to a lattice
RSOS
model with $t-1$ heights. In the massive case, the continum limit is
characterized by a quantum parameter $q_{massive}=\exp(i\pi/(t-1))$. The S
matrix
is based on a RSOS model with $t-2$ heights. The TBA diagram has $t-3$ heights
(
$t-4$ come from pseudoparticles, one is massive). In the massless case the
continuum limit has parameter $q_{massless}=\exp(i\pi/(t-2))$. The S matrix is
based on a RSOS model with $t-3$ heights. The TBA diagram has still $t-3$
heights
 ($t-5$ come from pseudoparticles, two massless). For instance it we
take $t=5$, $s=1$ that is the lattice $A_4$ model, in the  case of array I
we get the massive  flow from  the tricritical Ising theory
($c=7/10$) to  the trivial
 fixed point,
and in the case of array II the massless flow from tricritical Ising to
 Ising.

\newsec{Generalizations}

The method can be extended easily to more complicated arrays, or to other
 algebras.
 As an example let us focus on $so(2r)$. As is generic in the higher rank case,
the analysis of the Bethe ansatz equations, together with a string hypothesis
for every color $a=1,\ldots,r$ leads to a TBA system that can be associated
 with a diagram  in the plane order of roots versus color \BR\ . Restricting
to the case $q=\exp(i\pi/2r)$ and turning moreover to the RSOS case there are
only real solutions (1 strings) to take into account here, and the diagram is
 again a $D$ diagram, where this time the nodes are labelled by colors. We can
 then consider several kinds of arrays as before. We restrict now to
ferromagnetic interactions \foot{Recall that in level rank duality, ferro
 and antiferro interactions are exchanged \KR\ \ref\AS{D.Altschuler, H.Saleur,
 Nucl. Phys. B354 (1991) 579.}}. A simple case is the type I array built with
 the fundamental representation. Then one recovers the same results as in
 above (see discussion before eq \bet\ for $s=1$), due to the well known fact
 that the rational points of the sine Gordon model $\beta^2=8\pi{r\over r+1}$
can also be described as cosets $so(2r)_1\otimes so(2r)_1/so(2r)_2$ (a similar
 correspondence with the sine-Gordon model at  $\beta^2={8\pi\over r+1}$ would
 be observed in the antiferromagnetic case, where all nodes would be massive,
corresponding to the bound states).
 This correspondence  extends to higher spin. The most interesting case occurs
 with the spinor representations. Consider first the  array of type I based say
 on positive chirality. One
 gets then in the limit $\Lambda\rightarrow\infty$, thermodynamic
equations that coincide with the TBA associated with
 \bigskip
\noindent
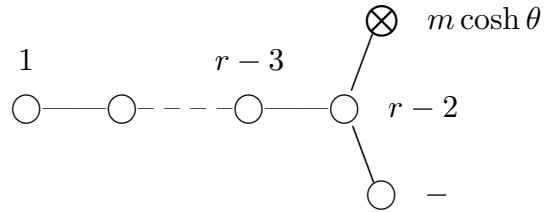
\centerline{\hbox{\rlap{\raise28pt\hbox{$\hskip4.5cm\bigotimes\hskip.25cm
m\cosh\theta$}}
\rlap{\lower27pt\hbox{$\hskip4.4cm\bigcirc\hskip.3cm -$}}
\rlap{\raise15pt\hbox{$\hskip4.1cm\Big/$}}
\rlap{\lower14pt\hbox{$\hskip4.0cm\Big\backslash$}}
\rlap{\raise15pt\hbox{$\hskip.2cm 1\hskip2cm r-3$}}
$\bigcirc$------$\bigcirc$-- -- --
--$\bigcirc$------$\bigcirc$ \hskip.2cm $r-2$}}

\bigskip
\noindent

This corresponds to the massive perturbation of $Z_r$ parafermions by the
 operator of dimension $h={r-1\over r}$ \ref\FZ{V.A.Fateev, Al.B.Zamolodchikov,
 Phys. Lett. B271 (1991) 91.}. For the array of type II, using the two
chiralities for $s_1$ and $s_2$ one finds in the limit $\Lambda\rightarrow
\infty$ thermodynamic equations that coincide with the TBA associated with

\bigskip
\noindent
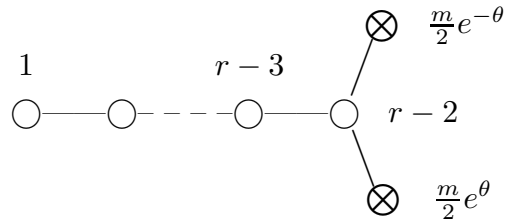
\centerline{\hbox{\rlap{\raise28pt\hbox{$\hskip4.5cm\bigotimes\hskip.25cm
{m\over 2}e^{-\theta}$}}
\rlap{\lower27pt\hbox{$\hskip4.4cm\bigotimes\hskip.3cm{m\over 2}e^\theta $}}
\rlap{\raise15pt\hbox{$\hskip4.1cm\Big/$}}
\rlap{\lower14pt\hbox{$\hskip4.0cm\Big\backslash$}}
\rlap{\raise15pt\hbox{$\hskip.2cm 1\hskip2cm r-3$}}
$\bigcirc$------$\bigcirc$-- -- --
--$\bigcirc$------$\bigcirc$ \hskip.2cm $r-2$}}

\bigskip
\noindent
 which corresponds to  the massless perturbation of parafermions
with $c_{UV}={2(r-1)\over r+1},\ c_{IR}=1-{6\over (r+1)(r+2)}$.
If we now consider the $r\rightarrow\infty$ limit, these two
 possibilites provide integrable  discretizations of the $O(3)$ non
linear sigma model at $\Theta=0$ (resp. $\Theta=\pi$) following \FZ\ .

Finally, it is  easy to read the S matrix for these two problems.
 For doing that we interpret the TBA diagram again as an $su(2)$ diagram.
 In the first case the $A_{r-1}$ part then arises from the pseudoparticles
 and the additional leg indicates that the physical particles have spin 2.
 Hence $S=S_{r-1}^{22}$ . In the second case we have to interpret the $A_{r-2}$
 part as related to pseudoparticles, and the two additional legs correspond
to the physical L and R particles. Then $S_{LL}=S_{RR}=S_{RL}=S_{r-2}^{11}$.

\newsec{Conclusion}

In conclusion we would like to notice that the approach where one
chooses for hamiltonian (eg in the type I array)
$$
H_I={1\over i}\ln\left[t^s(i\Lambda/2)(t^s(-i\Lambda/2))^{-1}\right]
$$
(as is done in \DV\ ) seems correct for the study of the scaling limit
but breaks down in the perturbative analysis \scconI\ . In that case instead
of equation \pert\ or \adI\ one gets a trivial unperturbed term, and the
identification with perturbed conformal field theories becomes questionable.
This problem is presumably due to the hamiltonian being non local.

We also would like to comment about the study of higher
conserved quantities from the lattice. For simplicity we restrict
to the massive case. Expand the energy \dI\ in the region
 $|\alpha|<\Lambda$, after renaming $e\rightarrow e^{(1)}$,
\eqn\highen{e^{(1)}=4\sum_{n=0}^{\infty}\exp\left[-(2n+1){t\Lambda\over 2}
\right]\cosh\left[(2n+1){t\alpha\over 2}
\right].}
Instead of \hami\ we can consider other hamiltonians obtained by
 taking higher derivatives. Define therefore
\eqn\hamoi{H_{I}^{(n)}=\left.i\left({i\over t}\right)^n{d^n\over du^n}\ln
\left[t^s(i\Lambda/2+u)(t^s(-i\Lambda/2-u))^{-1}\right]\right|_{u=0},}
Its eignvalues are given by
\eqn\hamoii{E_I^{(n)}=\left(-{2\over t}\right)^n\sum_kp_s^{(n)}
(\alpha_k+\Lambda)-(-1)^np_s^{(n)}(\alpha_k-\Lambda).}
The analysis of thermodynamics and excitations is very similar to the case
$n=1$.
 One finds the energies of excitations to be given by
\eqn\hamoiii{e^{(n)}(\alpha)=\left(-{2\over t}\right)^{n-1}\left({d\over
d\Lambda
}\right)^{n-1}e^{(1)}(\alpha).}
Expand now these quantities as \highen\ . One finds for instance
\eqn\hamoiv{e^{(2)}=4\sum_{n=0}^\infty (2n+1)
\exp\left[-(2n+1){t\Lambda\over 2}\right]\cosh\left[(2n+1){t\alpha\over
2}\right].}
All these quantities have dominant behaviour $e^{(n)}(\alpha)\approx
4e^{-t\Lambda/2}
\cosh(t\alpha/2)$. However by forming linear combinations of them we can
extract
higher order terms. For instance
\eqn\hamov{e^{(2)}-e^{(1)}\approx 8 e^{-3t\Lambda/2}\cosh(3t\alpha/2).}
Hence
\eqn\hamovi{H^{(2)}_{latt}-H^{(1)}_{latt}\approx N{\cal
E}_0+\Delta^3H_{scal}^{(3)},}
with obvious generalizations to higher orders. In the same way as
 $H_{scal}^{(1)}$ is  the quantum sine-Gordon  hamiltonian,
   the $H_{scal}^{(2n+1)}$  are the higher hamiltonians -
non trivial for odd grade only -  of the integrable hierarchy
 \ref\SY{R.Sasaki, L.Yamanaka, Adv. Studies in Pure Math. 16
(1988) 271.}
\EY\ . They commute since the various latttice hamiltonians \hamoi\  used
in their construction do.

\bigskip

\noindent{\bf Acknowledgments}: N.Yu Reshetikhin was
 supported by NSF under grant No. DMS-9015821 and the Sloane foundation.
H.Saleur
 by DOE under grant No. DE-FG03-84ER40168 and the Packard foundation.

\listrefs
\bye